\documentclass[%
reprint,
superscriptaddress,
amsmath,
amssymb,
pra,
]{revtex4-1}

\usepackage{xcolor}
\usepackage{braket}
\usepackage{hyperref}

\usepackage{graphicx}% Include figure files
\usepackage{dcolumn}% Align table columns on decimal point
\usepackage{bm}% bold math
\usepackage{array}
\begin{document}

\preprint{APS/123-QED}

\title{
%Broadband sensitive detection using Stark shifting
%for frequencies from 10~MHz to 1~GHz\\
%or\\
%Detection of 10~MHz to 1~GHz signals using Stark shifting in Rydberg-atom sensors\\ or \\
%Investigating the detection of 10~MHz to 1~GHz signals using Stark shifting in Rydberg-atom sensors\\ or\\
%Sensitive and Broadband Reception of Radio Frequency fields from 50 MHz to 1 GHz or\\
Characteristics of VHF and UHF band detection using Stark shifting in Rydberg-atom sensors
}% Force line breaks with \\

\author{Rajavardhan~Talashila} 
\affiliation{Department of Electrical Engineering, University of Colorado, Boulder, Colorado 80302, USA}
\affiliation{Associate of the National Institute of Standards and Technology, Boulder, Colorado 80305, USA}
\author{Noah~Schlossberger}
\affiliation{ National Institute of Standards and Technology, Boulder, Colorado 80305, USA}
\author{David~S.~La Mantia}
\author{Abby~Halasi-Kun}
\author{Baran~Kayim}
\author{Michael~A.~Viray}
\author{Robert~Wyllie}
\affiliation{Georgia Tech Research Institute, Atlanta, Georgia 30332, USA}
\author{Christopher L. Holloway}
\author{Nikunjkumar Prajapati}
\email{Corresponding Author: nikunjkumar.prajapati@nist.gov}
\affiliation{ National Institute of Standards and Technology, Boulder, Colorado 80305, USA}

\date{\today}% It is always \today, today,

\begin{abstract}
We present a characterization of the sensitivity of broadband electric field measurements in the Very High Frequency (VHF) and Ultra High Frequency (UHF) bands using AC stark shifting of Rydberg states. The dependence of the sensitivity and dynamic range of the field measurements using $nD_{5/2}$ states of cesium for principal quantum numbers n= 60, 70, 80, 90 and 100 is presented, detailing the intricacies involved. The frequencies considered for the measurement range from 50~MHz to 900~MHz for which a transverse electromagnetic (TEM) cell is used to excite the Rydberg atoms with the RF fields. The relative sensitivities achieved for various principal quantum numbers are presented. A general trend of better sensitivity for higher principal quantum numbers is observed. This work paves the way for a better understanding of the AC Stark dependent field measurements using Rydberg sensors.
\end{abstract}

\maketitle

%\tableofcontents

\section{Introduction}
Classical radio-frequency (RF) receivers require matched antennas for reception across different bands, and typically require impedance-matching circuits, frequency mixers, and various-sized antennas to bridge frequency gaps in reception~\cite{stutzman_antenna_2013}.  
Thus, the reception of radio waves in a single device that can range from MHz to GHz would require several antennas, various electronics, and many filters for sensitive field reception~\cite{raab_hf_2002,zhang_propagation_2009,mattioni_design_2007}. 
Rydberg atom electrometers have promise for overcoming this barrier and reducing complexity. These sensors have demonstrated promise for DC and AC field detection with significant potential for broadband operation \cite{holloway_broadband_2014,holloway_electromagnetically_2022, sedlacek_microwave_2012}. 

Rydberg atoms have been utilized for electric field detection for  well over a  decade. 
The first demonstrations mapped out traceability to the international system of units as the primary driver for development ~\cite{holloway_electric_2017, sedlacek_atom-based_2013}. 
The use of Rydberg sensors has expanded to communications reception, temperature sensors, angle-of-arrival, radar, and many other applications have emerged ~\cite{schlossberger_rydberg_2025, yang_pcb-based_2024, allinson_simultaneous_2024,chen_electric_2025,paradis_atomic_2019, chen_electric_2025, yang_local_2025, prajapati_high_2024, glick_low-frequency_2026}. Recent work has utilized Rydberg atomic sensors to detect
 high frequency (HF band: 3 MHz to 30 MHz), very high frequency (VHF band: 30 MHz to 700 MHz), and ultra high frequency (UHF band: 700 MHz to 3 GHZ) bands.
In the realm of frequencies less than 1 GHz, various teams have shown new methods of detection for these fields and have presented competitive sensitivities via multi-photon approaches or coupling to higher angular momentum states via RF coupling \cite{song_continuous_2024, yang_local_2025, PhysRevA.107.052605, richardson_extraction_2025, kayim_calibration_2026}. 

\begin{figure}
    \centering
    \includegraphics[width=0.99\linewidth]{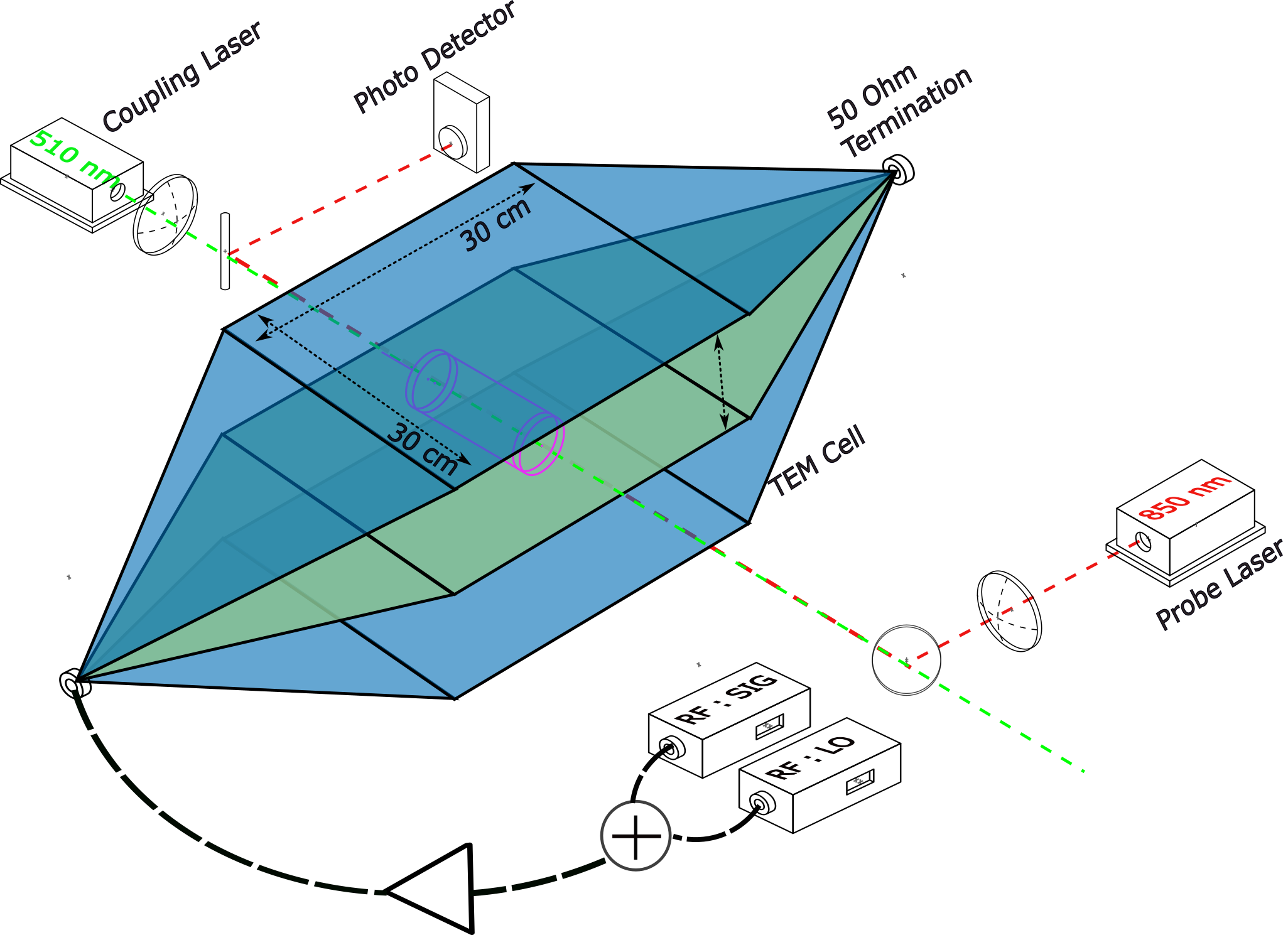}
    \caption{Experimental Setup with Cs vapor cell placed within the TEM cell connected to RF sources through an amplifier. The cylindrical cell has diameter of 25 mm and length 75 mm. The TEM cell has the central septum separated with the outer plate by 10 cm  and the rectangular portion of the parallel plates are 30x30 cm in dimension. The field the Rydberg atoms experience is calibrated by using the AC Stark shifting method.}
    \label{fig:Experimental_Setup}
\end{figure}

In this manuscript, we revisit the simple case of two-photon electromagnetically induced transparency (EIT) in cesium. 
We avoid Autler-Townes splitting since the phenomenon is largely limited to resonant electric fields with the exception of  Rydberg engineering for modifying energy levels~\cite{noauthor_rydberg-state_nodate}. 
We instead utilize the AC Stark shift to map out the response of the Rydberg atoms across various frequency bands and principal quantum numbers (n). 
Specifically, we characterize the changes in sensitivity, shielding, and resonances in a transverse-electromagnetic (TEM) waveguide using the atoms as sensors, as shown in Fig. \ref{fig:Experimental_Setup}.
We theoretically map out the atomic and TEM cell response and attribute these characteristics to atomic states, electric field shielding, and atomic resonances using electric field modeling techniques and the ARC Rydberg calculator \cite{sibalic_arc_2017}.

\begin{figure}
    \centering
    \includegraphics[width=0.65\linewidth]{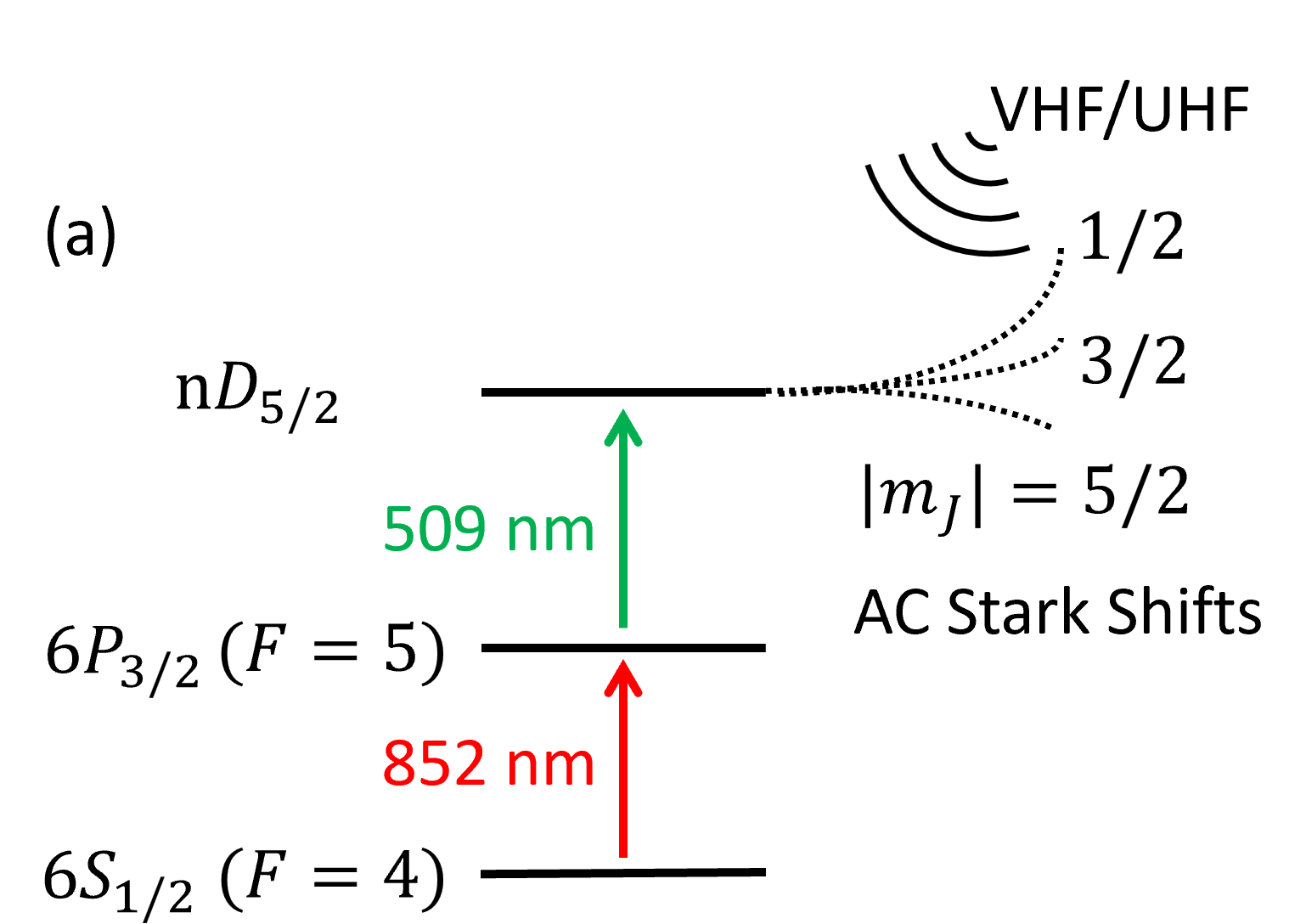}
    \includegraphics[width=0.99\linewidth]{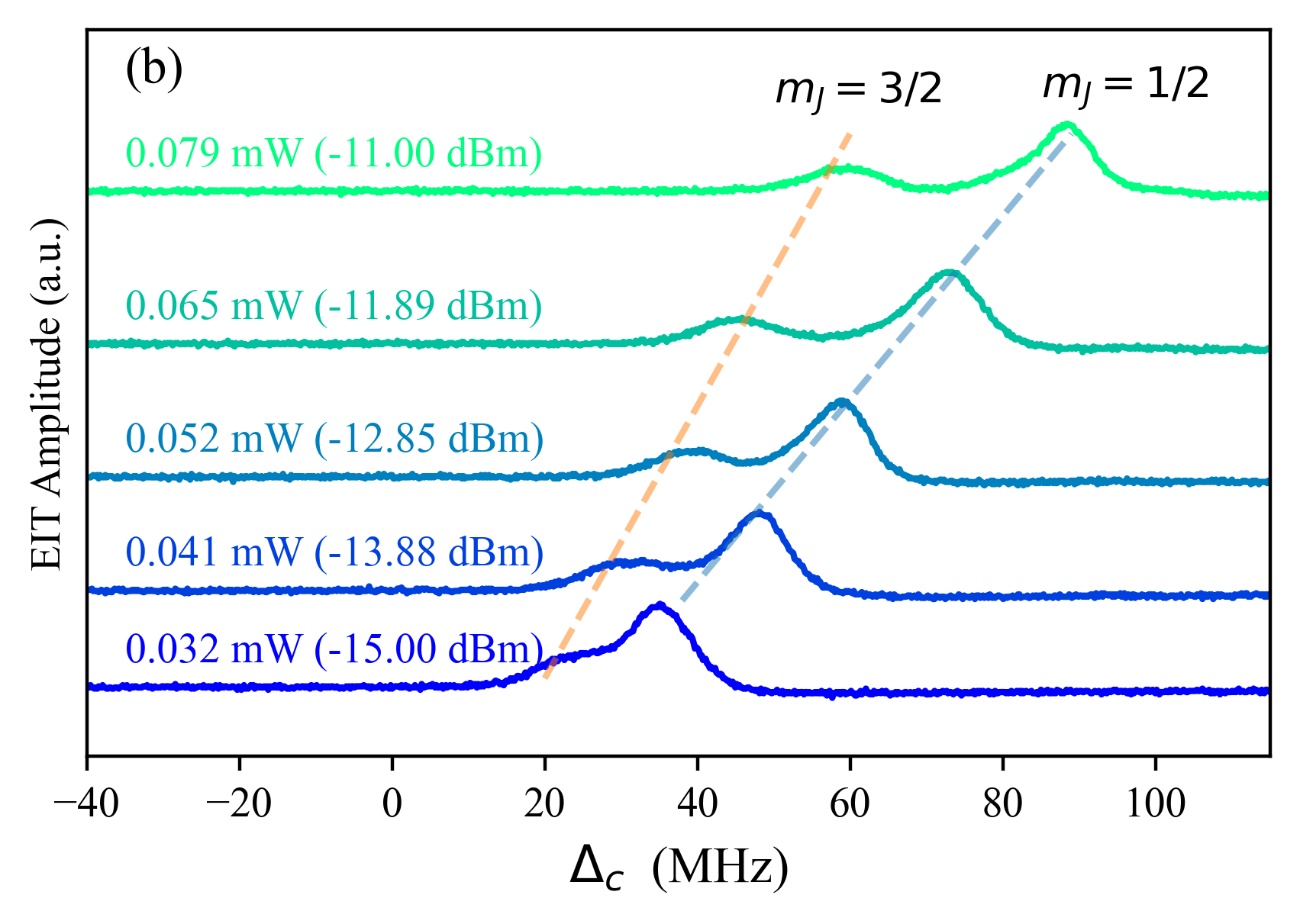}
    \caption{(a) Level diagram. The states $6S_{1/2}$, $6P_{3/2}$ and the Rydberg state $nD_{5/2}$ are resonantly driven by the probe (852 nm) and coupling (509 nm) lasers for obtaining the EIT. The applied RF field is responsible for the AC stark shifting of the EIT spectrum. (b) The obtained AC stark shifted EIT spectrum for the applied RF field of 200 MHz at increasing power levels. The $m_J$ peaks are fitted to  the expected Stark shifts using eq.\ref{eq:Stark}. The calibration factor obtained across the VHF and UHF bands is shown in Fig.~\ref{fig:Calibration}.  }
    \label{fig:Level_Diagram}
\end{figure}

\section{Stark based Field Measurements and Calibration of the Field}
\begin{figure}[h]
    \centering
    \includegraphics[width=0.99\linewidth]{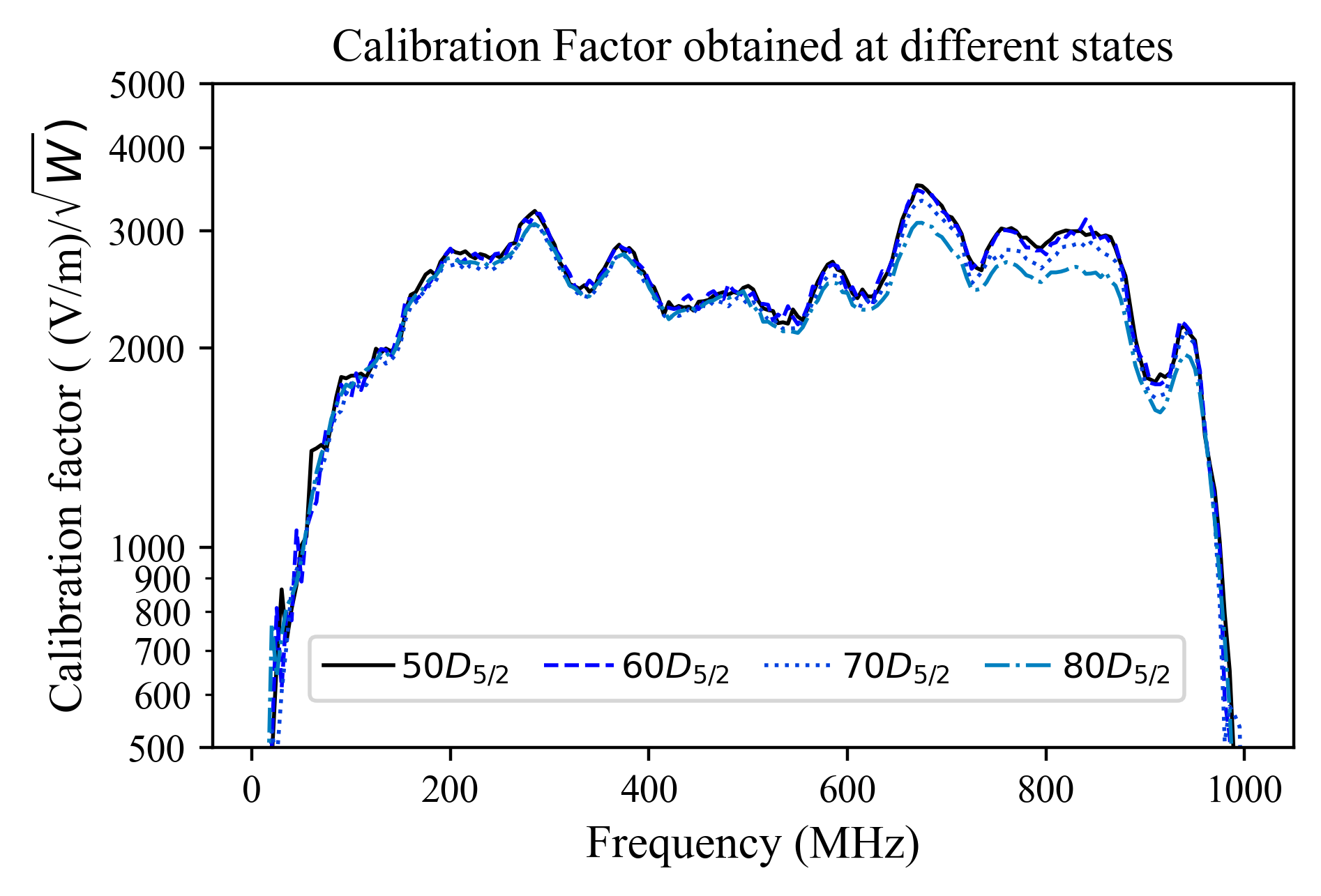}
    \caption{The calibration factor is obtained from fitting the Stark shifting for different states. They all overlap as expected. The TEM cell has a good range from around 50 MHz to 900 MHz. The deviation for the case of the $80D_{5/2}$ state at higher frequencies arises from the influence of adjacent states, so the behavior is not a purely Stark-like shift.\cite{holloway_broadband_2014}  }
    \label{fig:Calibration}
\end{figure}
We utilize electromagnetically induced transparency, a two-photon excitation scheme that allows for Doppler-reduced measurements to probe the Rydberg states of interest (n=60,70,80,90,100), shown in Fig.~\ref{fig:Level_Diagram}(a).
We utilize a two-photon scheme to excite the cesium atoms to the Rydberg state.
The nature of this laser allows for broad tunability across various principal quantum numbers. 
These lasers are passed through a 75 mm long by 25 mm diameter vapor cell in a counter-propagating orientation for optimal Doppler selection.
The coupling and probe beams have approximate Full Width at Half Maximum (FWHM) diameter of 280 $\mu$m and the powers are fixed to 56 mW and 18 $\mu$W respectively, for this study. 
Both lasers were locked for measurements using an ultra low expansion cavity with offset Pound-Drever Hall locks. 
Finally, readout of the Rydberg states was done by measuring the probe laser on a photodetector. 

Broadband reception of fields also requires broadband generation of fields. 
Standard antennas do not have the range needed to generate uniform fields with reliable and stable powers and polarizations across various frequencies.
To this end, we utilized a transverse electromagnetic (TEM) cell to generate stable and uniform fields from DC to $1\,\mathrm{GHz}$, shown in Fig.~\ref{fig:Experimental_Setup}.
Beyond this, even the TEM cells begin to fail in providing a reliable output and thus limited the scope of this experiment.
The vapor cell of interest was placed in the center of open-sided TEM cell and simulations for expected field patterns with frequency dependence were performed to verify operation.
We also performed measurements of the 2-port S-parameters of the cavity with a vector network analyzer to determine multi-mode resonances and did not observe any in the range 50 MHz to 900 MHz.

We applied and measured RF fields within the standard operating range of the experimental setup. This measurement utilized the Stark effect, shown in Fig.~\ref{fig:Level_Diagram}(b) \cite{steck_quantum_nodate, holloway_electromagnetically_2022}
    \begin{equation}\label{eq:Stark}
        \Delta f_{stark} = - \cfrac{1}{2} \left( \alpha_0 + \alpha_2 \cfrac{3m_J^2 - J(J+1)}{J(2J -1)}  \right) E^2 = -\cfrac{1}{2} \alpha E^2
    \end{equation}
where $\alpha_0$ and $\alpha_2$ correspond to scalar and tensor polarizabilities, respectively. Here, $\Delta f$ corresponds to the AC Stark shift and $E$ is the rms value of the electric field. 
The vector polarizability $\alpha_1$ does not contribute to the Stark Shifting for the case of linear polarization of the applied RF field which is expected in the TEM cell \cite{steck_quantum_nodate}. 

By measuring the shift of the atomic states and fitting, we obtained a value for the actual field measured at the location of the atoms for the different frequencies of interest.
However, we could also compare this value to a simulation for the expected field or estimates for field strength in a TEM cell based on its dimensions. 
The discrepancy between the field expected for a given power and the field measured is referred to as a calibration factor.
This factor can be attributed to different contributions like electric field shielding from the vapor cell and resonances depending on the frequency being probed and is elaborated upon in the discussion.
However, the factor is wholly separated from any atomic properties and therefore should be consistent across various Rydberg states.
We measure the shifts over a range of RF powers applied to increase statistics and verify linear scaling of our RF power across all powers considered using a spectrum analyzer.

The calibration factor in our system is

\begin{equation}
    \text{Calibration Factor} = \cfrac{E_{\text{Measured}}}{\sqrt{P_{\text{in}}}}= -\cfrac{1}{\alpha}\cfrac{\sqrt{2\Delta f}}{\sqrt{P_{\text{in}}}},
\end{equation}
where $P_{\text{in}}$ is applied RF power from the signal generator, $E_{\text{Measured}}$ is the measured Electric field magnitude, $\Delta f$ is the measured shift of the $m_J$ peak in the spectra and $\alpha$ is the polarizability as described in Eq.\ref{eq:Stark}.

This calibration factor was obtained for frequencies ranging from 50 MHz to 1 GHz for different principal quantum numbers (n=60,70,80,90).
We observed good agreement across different $n$, with the exception of states and frequencies where resonant transitions come into play and disrupt the fitting, shown in Fig.~\ref{fig:Calibration}.
Slight deviations ($<$ 5\%) can be seen for these regions and the primary effect of these resonances is on the optimization of the local oscillator (LO) field providing the biasing and reference for measuring the signal field of interest.
However, we can attribute these deviations to nonconventional polarizability response from additional nearby atomic states. For the cases of $n=90$ and $n=100$, due to the atomic resonant frequencies below 1 GHz, the EIT spectrum undergoes AT splitting and hence are not considered for calibration using the AC Stark shifting method.

\section{Atom Heterodyne Optimization and Sensitivity}
\begin{figure}
    \centering
    \includegraphics[width=0.99\linewidth]{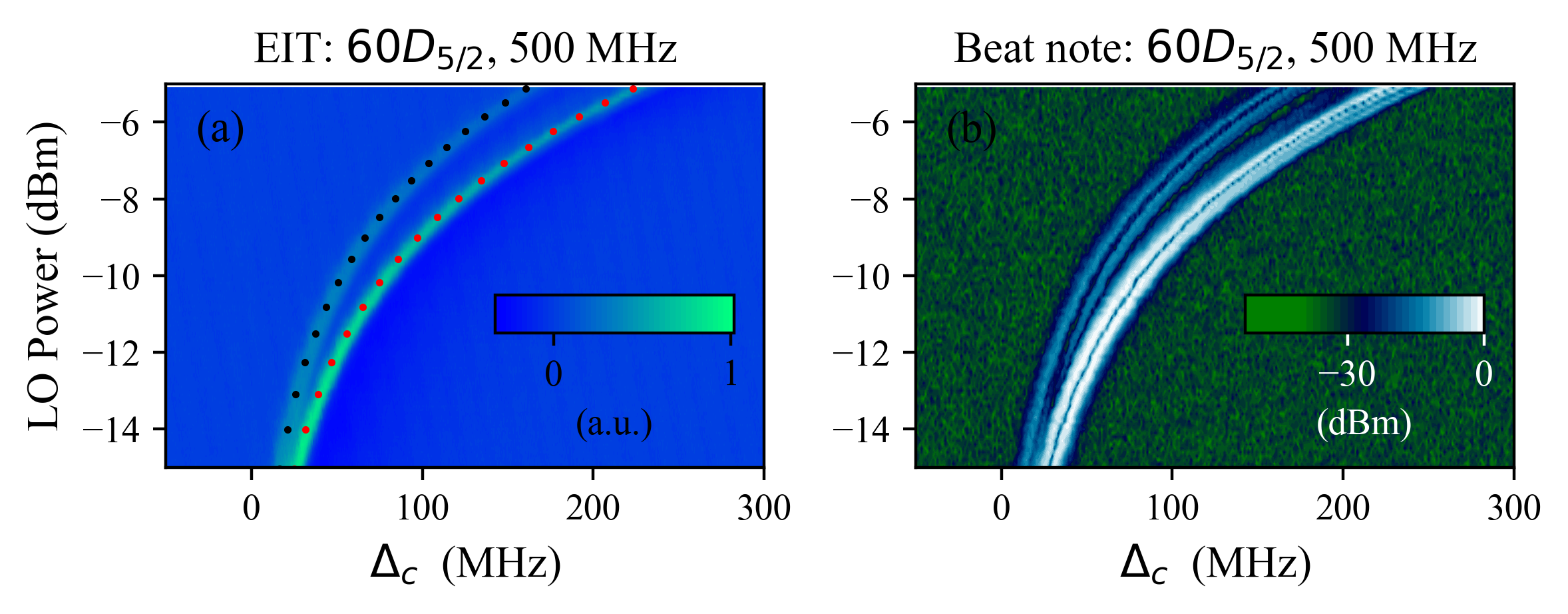}
    \caption{(a)  EIT spectra under AC stark shifting for the states $60D_{5/2}$ applied RF frequency of 200 MHz. The  x-axis corresponds to coupling laser detuning and y-axis is applied LO power in dBm. Calculated Floquet spectra using the Shirley method of ARC \cite{sibalic_arc_2017} are overlaid as red and black dots for $m_J= 1/2$ and $3/2$ cases, respectively.  (b) The signal power was maintained at a constant reference level with a 200 kHz offset} from the LO frequency. The amplitude of the beat note  is captured in the same 2D space by the  spectrum analyzer. LO power and  coupling detuning for maximum beat note amplitude are extracted from this exercise and are set to this optimal point for extracting the sensitivity.
    \label{fig:Beatnote_Optimize}
\end{figure}
Rydberg atoms have demonstrated phase sensitivity in the presence of a reference field or LO \cite{jing_atomic_2020,simons_rydberg_2019}.
The introduction of an LO offset in frequency results in the generation of a beat note that can be measured on a spectrum analyzer with significantly improved signal to noise ratio (SNR), and is the basis for all communication protocols.
Similar to communication protocols, there is an optimal LO power that will give the best SNR defined by the mixer used.
The atoms have a similar operation where the scaling of the AC Stark shift results in an effective amplification of the signal as the LO is increased

\begin{align}
    \Delta f_{stark} \approx  -\cfrac{1}{2} \alpha \left[ E^2_{LO} +E^2_{sig} + 2 E_{LO} E_{sig} \sin(\Delta_{beat} t)  \right],
\end{align}

where $E_{LO}$  and $E_{sig}$ are the field strength of the LO and signal respectively. However, in the case of the atoms, there is an optimal LO as increasing the applied field leads to broadening of the special feature, limiting this gain behavior, shown in Fig.~\ref{fig:Beatnote_Optimize}(a) and (b). 
The data shows that as the RF power applied increases and the peak shifts in frequency, the beat note amplitude initially increases according to the scaling in eq.\ref{eq:Stark}, but then begins to decrease. 
By employing opposing standing waves to reduce the non-uniformity of the fields, we may be able to push this technique even further and improve the sensitivity~\cite{rotunno_investigating_2023}.
However, this is outside of the scope of this exploration. 

\begin{figure}
    \centering
    \includegraphics[width=0.99\linewidth]{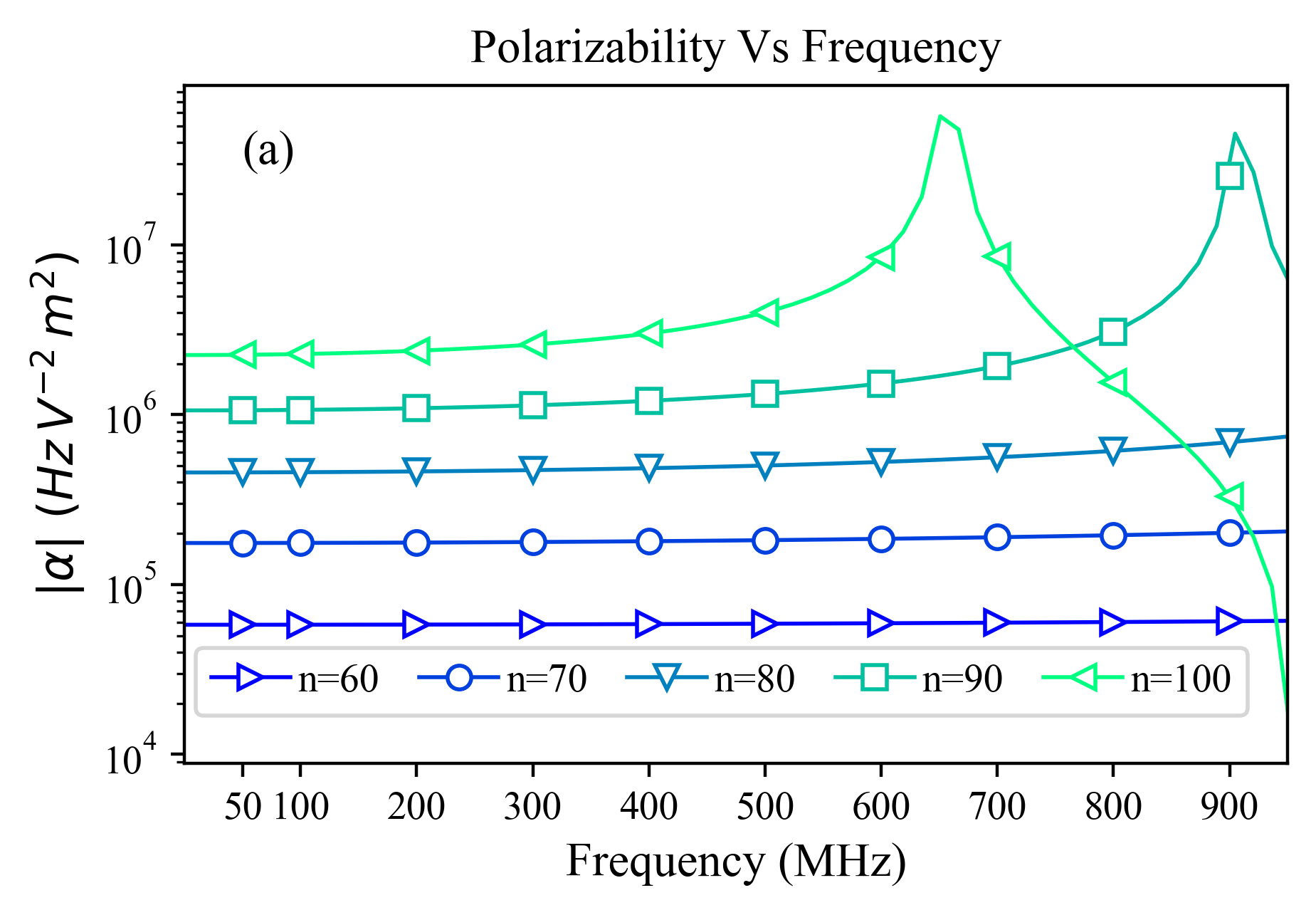}
    \includegraphics[width=0.99\linewidth]{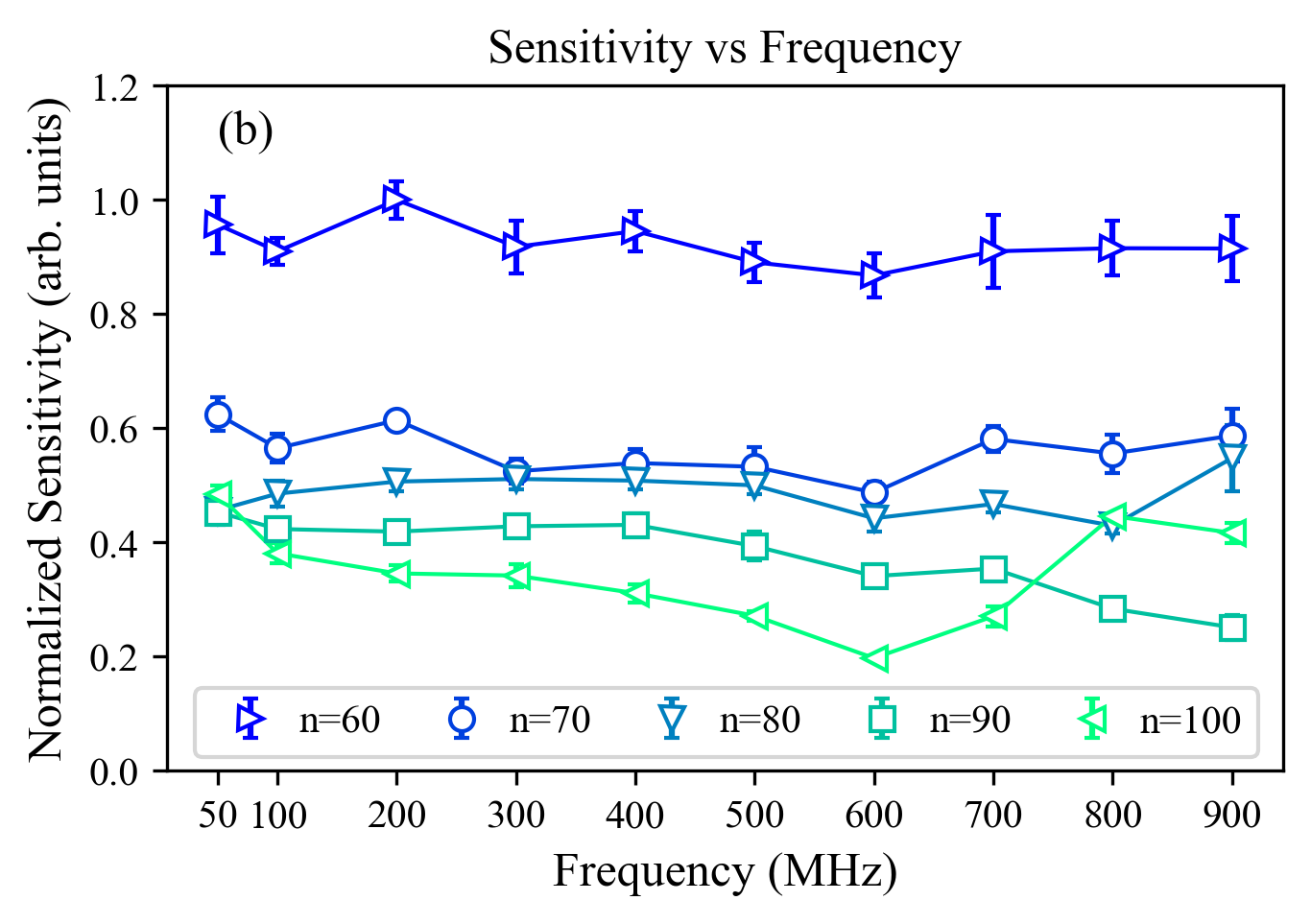}
    \caption{(a) Calculated values of the magnitude of the  polarizabilities plotted against frequency for various states $nD_{5/2}$ for $m_J=1/2$ using ARC \cite{sibalic_arc_2017} . For higher principal quantum numbers, the observed atomic resonant transitions are consistent with theoretical models. (b) Measured sensitivities plotted against LO frequencies associated with AC Stark shifts of the $nD_{5/2}$ state in the $n=60$ to $n=100$  range.}
    \label{fig:Sensitivities_Measured}
\end{figure}

Sensitivity measurements for RF frequencies in the range of 50 MHz to 900 MHz are obtained for the states of $nD_{5/2}$ in the range n=60 to 100. The beat note frequency, i.e the difference in the signal and LO frequencies is $\Delta_{beat} = 200$~kHz. This beat note frequency is chosen as it is past the $1/f$ noise range of the photodetector spectrum. For each measurement, maintaining the signal power within the linear operating regime,  the beat note amplitude is maximized in the 2D space of applied LO power and coupling laser detuning as demonstrated in Fig.~\ref{fig:Beatnote_Optimize}. From this state of the coupling laser, in the absence of the signal power, the noise floor is recorded. In the next measurement step, beat note amplitude as a function of the applied signal power is measured. The obtained data is analyzed to obtain its intersection with the noise floor and hence deduce the sensitivity. The spectrum analyzer is set to a resolution bandwidth of 1 kHz in these measurements. The sensitivity values obtained by the above description are plotted in Fig.~\ref{fig:Sensitivities_Measured}(b). Each measurement is repeated 5 times and the corresponding standard deviations are plotted as error bars. 

\begin{figure}
    \centering
    \includegraphics[width=0.99\linewidth]{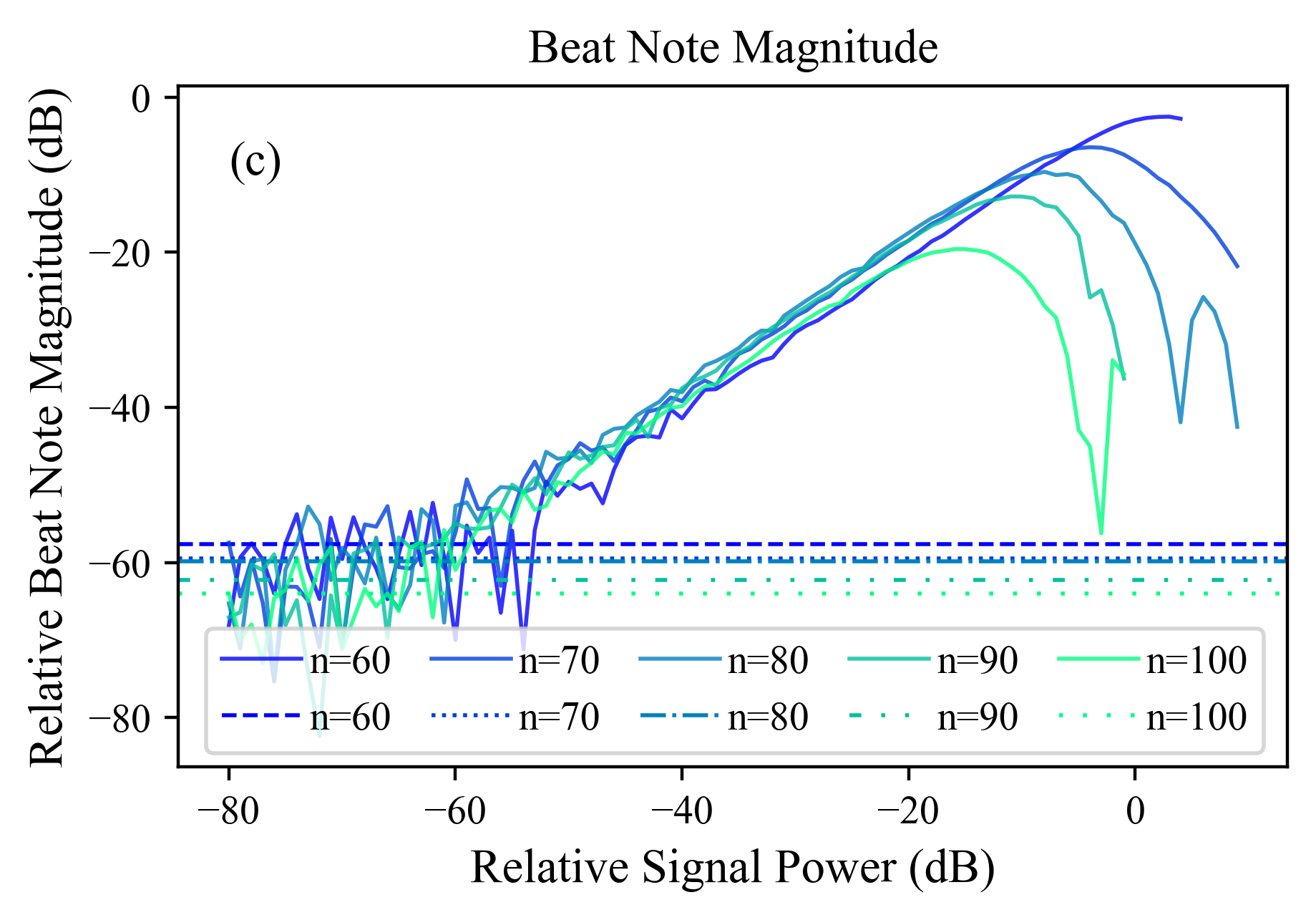}
    \caption{The beat note magnitude against the sweep of the signal power for various states $nD_{5/2}$ for the case of $\omega_{LO}=200$ MHz. The corresponding noise floors are also plotted. The shot noise corresponding to the probe power of 5.1 $\mu$W reaching the photodetector is $\approx -84\,\mathrm{dBm}$. }
    \label{fig:Dynamic_Range}
\end{figure}

The beat note optimization sets the amplitude of the LO applied for a given RF frequency. We determine the sensitivity following the work in~\cite{prajapati_enhancement_2021,holloway_broadband_2014}. We fix the LO power and obtain traces on a spectrum analyzer as the power of the signal RF is decreased into the noise floor. The intercept of the  slope created by the beat note amplitude and the noise floor gives the minimum detectable field ($E_{min}$). The sensitivity is then given by the $E_{min}/\sqrt{RBW}$ where RBW is the resolution bandwidth or collection time. Sample data for $\omega_{LO}=200$ MHz for various $n$ along with the corresponding noise floors are shown in Fig.~\ref{fig:Dynamic_Range}.

We sweep the signal powers to fully cover the lower range where the beatnote signal is under the noise floor and to the point where we can see a nonlinear response. This shows the dynamic range of the atomic system. The lower we tune in principal quantum number, we observe that the dynamic range of the system can be improved and is nearly 60 dB. For the case of the higher principal quantum number, the range is reduced to 40 dB. This is due to the large polarizability of the higher principal quantum number states. As the signal field is increased, we observe a large AC stark shift at lower field strengths. This is also reflected by the fact that the optimal LO for the different states decreases significantly for higher principal quantum number.

\section{Discussion\label{sec:Discussion}}
We have demonstrated several important aspects in this work. The most important of which is the broadband sensing and calibration capabilities of the Rydberg atom sensor. In this work, we demonstrate highly sensitive broadband field detection across the targeted VHF and UHF bands. While these values represent local sensitivities, they highlight the potential of Rydberg atoms for sensitive RF detection.

Looking to the graph of polarizability in Fig.~\ref{fig:Sensitivities_Measured}(a), a flat response is observed from DC until the first resonance with a nearby Rydberg state. The spike in the polarizability show the locations of these atomic resonances. Typically, the high frequency side of the RF resonance experiences a sharp drop in the polarizability, which in turn diminishes the sensitivity of the device. Also, as you go high in principle quantum number, the nearby states come increasingly closer in energy and alter sensor operation. We have demonstrated sensitivity measurements for 5 different states, shown in Fig.~\ref{fig:Sensitivities_Measured} (b). Here we can see clearly, a mostly flat response across the entire measurement range with slight deviations when looking at n=100. In this case, the sharp spike in the polarizability and some avoid crossing effects \cite{talashila_floquet_2026} play a strong roll in changing the response of the Rydberg atoms sensor. However, even with the resonances present, we have observed the best sensitivities at this principle quantum number.

The fact that n=100 shows the best sensitivity is likely due to its substantially higher polarizability. In our experiment, we fixed the power of the coupling laser. This would typically result in decreased EIT signal height as the principle quantum number is increased. However, the EIT signal strength scales as $n^2$ while the polarizability scales as $n^7$ . This allows for the overall increased sensitivity. While we were limited in total laser power, we were far from saturating the transition. There may have been room for improvement if more laser power was available to compensate and maintain similar Rabi rates for the coupling laser.

Figure \ref{fig:Dynamic_Range} presents of the characteristics of the beat note amplitude versus the signal power sweep for various states and the corresponding noise floors. The noise floor drops by few dB with increasing $n$, however, the saturation point of the signal power drops steeply. Hence, the overall dynamic range drops with increasing $n$. This will have implications on the range of signal powers the Rydberg sensor will be responding to. This behavior calls for further investigation about the noise floor behavior and resulting dynamic range for different $n$.

These demonstrations open the door to building portable broadband sensors based on Rydberg atom work. Typical use of Autler-Townes splitting requires a dynamic and agile laser system that does not exist yet to quickly tune between different atomic states to sense different RF frequencies. Prior efforts always claimed significantly worse sensitivities in the Stark measurement regime. This work closes that gap and provides a reference for ideal working conditions of the Rydberg atom sensor for broadband detection in the HF, VHF, and UHF bands.

\section{Conclusion}

In this manuscript, we have demonstrated the measurements of the sensitivity of RF field detection in VHF and UHF bands using Stark shifting. The measurements are conducted in the range of states $nD_{5/2}$ for $n=$ 60 to 100 and the fields coupled into the vapor cell are calibrated using AC Stark shifting of the EIT spectra. We found that though higher $n$ offers better sensitivities, the dynamic range of the beat note measurement drops. We also show that states with higher $n$ are riddled with closely spaced resonant transitions and possible avoided crossings \cite{talashila_floquet_2026}; hence, calling for explicit coupling laser lockpoints to overcome them in the RF heterodyne detection scheme.

\section{Acknowledgments}
A contribution of the U.S. government, this work is not subject to copyright in the U.S.
\bibliography{Refs}% Produces the bibliography via BibTeX.

@article{sibalic_arc_2017,
    title = {{ARC}: {An} open-source library for calculating properties of alkali {Rydberg} atoms},
    volume = {220},
    issn = {0010-4655},
    shorttitle = {{ARC}},
    url = {https://www.sciencedirect.com/science/article/pii/S0010465517301972},
    doi = {10.1016/j.cpc.2017.06.015},
    urldate = {2025-11-13},
    journal = {Computer Physics Communications},
    author = {Šibalić, N. and Pritchard, J. D. and Adams, C. S. and Weatherill, K. J.},
    month = nov,
    year = {2017},
    pages = {319--331},
}

@article{prajapati_high_2024,
    title = {High angular momentum coupling for enhanced {Rydberg}-atom sensing in the very-high frequency band},
    volume = {135},
    issn = {0021-8979},
    url = {https://doi.org/10.1063/5.0179496},
    doi = {10.1063/5.0179496},
    number = {7},
    urldate = {2024-11-20},
    journal = {Journal of Applied Physics},
    author = {Prajapati, Nikunjkumar and Kunzler, Jakob W. and Artusio-Glimpse, Alexandra B. and Rotunno, Andrew P. and Berweger, Samuel and Simons, Matthew T. and Holloway, Christopher L. and Gardner, Chad M. and Mcbeth, Michael S. and Younts, Robert A.},
    month = feb,
    year = {2024},
    pages = {074402},
}

@article{holloway_electromagnetically_2022,
    title = {Electromagnetically induced transparency based {Rydberg}-atom sensor for traceable voltage measurements},
    volume = {4},
    issn = {2639-0213},
    url = {https://pubs.aip.org/aqs/article/4/3/034401/2835268/Electromagnetically-induced-transparency-based},
    doi = {10.1116/5.0097746},
    number = {3},
    urldate = {2024-09-11},
    journal = {AVS Quantum Science},
    author = {Holloway, Christopher L. and Prajapati, Nikunjkumar and Sherman, Jeffery A. and Rüfenacht, Alain and Artusio-Glimpse, Alexandra B. and Simons, Matthew T. and Robinson, Amy K. and La Mantia, David S. and Norrgard, Eric B.},
    month = sep,
    year = {2022},
    pages = {034401},
}

@article{holloway_broadband_2014,
    title = {Broadband {Rydberg} {Atom}-{Based} {Electric}-{Field} {Probe} for {SI}-{Traceable}, {Self}-{Calibrated} {Measurements}},
    volume = {62},
    issn = {1558-2221},
    url = {https://ieeexplore.ieee.org/document/6910267},
    doi = {10.1109/TAP.2014.2360208},
    number = {12},
    urldate = {2024-09-11},
    journal = {IEEE Transactions on Antennas and Propagation},
    author = {Holloway, Christopher L. and Gordon, Joshua A. and Jefferts, Steven and Schwarzkopf, Andrew and Anderson, David A. and Miller, Stephanie A. and Thaicharoen, Nithiwadee and Raithel, Georg},
    month = dec,
    year = {2014},
    note = {Conference Name: IEEE Transactions on Antennas and Propagation},
    pages = {6169--6182},
}

@article{jing_atomic_2020,
    title = {Atomic superheterodyne receiver based on microwave-dressed {Rydberg} spectroscopy},
    volume = {16},
    copyright = {2020 The Author(s), under exclusive licence to Springer Nature Limited},
    issn = {1745-2481},
    url = {https://www.nature.com/articles/s41567-020-0918-5},
    doi = {10.1038/s41567-020-0918-5},
    number = {9},
    urldate = {2024-09-10},
    journal = {Nature Physics},
    author = {Jing, Mingyong and Hu, Ying and Ma, Jie and Zhang, Hao and Zhang, Linjie and Xiao, Liantuan and Jia, Suotang},
    month = sep,
    year = {2020},
    note = {Publisher: Nature Publishing Group},
    pages = {911--915},
}

@article{simons_rydberg_2019,
    title = {A {Rydberg} atom-based mixer: {Measuring} the phase of a radio frequency wave},
    volume = {114},
    issn = {0003-6951, 1077-3118},
    shorttitle = {A {Rydberg} atom-based mixer},
    url = {https://pubs.aip.org/apl/article/114/11/114101/36735/A-Rydberg-atom-based-mixer-Measuring-the-phase-of},
    doi = {10.1063/1.5088821},
    number = {11},
    urldate = {2024-08-13},
    journal = {Applied Physics Letters},
    author = {Simons, Matthew T. and Haddab, Abdulaziz H. and Gordon, Joshua A. and Holloway, Christopher L.},
    month = mar,
    year = {2019},
    pages = {114101},
}

@article{yang_local_2025,
    title = {Local {Oscillator} {Port}-{Integrated} {Resonators} for {Sensitivity} {Enhancement} of {VHF} {Band} {Rydberg} {Atomic} {Heterodyne} {Receivers}},
    issn = {1557-9670},
    url = {https://ieeexplore.ieee.org/document/10849580/},
    doi = {10.1109/TMTT.2025.3528720},
    urldate = {2025-08-06},
    journal = {IEEE Transactions on Microwave Theory and Techniques},
    author = {Yang, Kai and Lin, Yi and Ding, Zhenke and An, Qiang and Sun, Zhanshan and Fu, Yunqi},
    year = {2025},
    pages = {1--10},
}

@article{PhysRevA.107.052605,
  title = {Very-high- and ultrahigh-frequency electric-field detection using high angular momentum Rydberg states},
  author = {Brown, Roger C. and Kayim, Baran and Viray, Michael A. and Perry, Abigail R. and Sawyer, Brian C. and Wyllie, Robert},
  journal = {Phys. Rev. A},
  volume = {107},
  issue = {5},
  pages = {052605},
  numpages = {8},
  year = {2023},
  month = {May},
  publisher = {American Physical Society},
  doi = {10.1103/PhysRevA.107.052605},
  url = {https://link.aps.org/doi/10.1103/PhysRevA.107.052605}
}

@article{chen_electric_2025,
    title = {Electric field sensing via {Rydberg} electromagnetically induced transparency using {Zeeman} and {Stark} effects},
    volume = {138},
    issn = {0021-8979},
    url = {https://doi.org/10.1063/5.0277790},
    doi = {10.1063/5.0277790},
    number = {9},
    urldate = {2026-02-02},
    journal = {Journal of Applied Physics},
    author = {Chen, Yu-Chi and Fang, Shao-Cheng and Su, Hsuan-Jui and Chen, Yi-Hsin},
    month = sep,
    year = {2025},
    pages = {094501},
}

@article{paradis_atomic_2019,
    title = {Atomic measurements of high-intensity {VHF}-band radio-frequency fields with a {Rydberg} vapor-cell detector},
    volume = {100},
    url = {https://link.aps.org/doi/10.1103/PhysRevA.100.013420},
    doi = {10.1103/PhysRevA.100.013420},
    number = {1},
    urldate = {2026-02-02},
    journal = {Physical Review A},
    publisher = {American Physical Society},
    author = {Paradis, Eric and Raithel, Georg and Anderson, David A.},
    month = jul,
    year = {2019},
    pages = {013420},
}

@article{song_continuous_2024,
    title = {Continuous broadband {Rydberg} receiver using {AC} {Stark} shifts and {Floquet} states},
    volume = {125},
    issn = {0003-6951},
    url = {https://doi.org/10.1063/5.0227250},
    doi = {10.1063/5.0227250},
    number = {19},
    urldate = {2026-02-03},
    journal = {Applied Physics Letters},
    author = {Song, Danni and Jiao, Yuechun and Hu, Jinlian and Yin, Yuwen and Li, Zhenhua and He, Yunhui and Bai, Jingxu and Zhao, Jianming and Jia, Suotang},
    month = nov,
    year = {2024},
    pages = {194001},
}

@article{schlossberger_rydberg_2025,
  title = {Rydberg atom reception of a handheld UHF frequency-modulated two-way radio},
  author = {Schlossberger, Noah and McDonald, Tate and Prajapati, Nikunjkumar and Holloway, Christopher L.},
  journal = {Phys. Rev. Appl.},
  volume = {25},
  issue = {2},
  pages = {024031},
  numpages = {8},
  year = {2026},
  month = {Feb},
  publisher = {American Physical Society},
  doi = {10.1103/jlrg-6889},
  url = {https://link.aps.org/doi/10.1103/jlrg-6889}
}

@article{yang_pcb-based_2024,
    title = {{PCB}-{Based} {Electrically} {Tunable} {Resonator} for {VHF} {Band} {Rydberg} {Atomic} {Enhancement} {Sensing}},
    volume = {72},
    issn = {1558-2221},
    url = {https://ieeexplore.ieee.org/document/10499207/},
    doi = {10.1109/TAP.2024.3385918},
    number = {7},
    urldate = {2026-02-03},
    journal = {IEEE Transactions on Antennas and Propagation},
    author = {Yang, Kai and Chen, Xiaozhang and Mao, Ruiqi and Li, Jianbing and Fu, Yunqi},
    month = jul,
    year = {2024},
    pages = {6060--6068},
}

@article{allinson_simultaneous_2024,
    title = {Simultaneous multiband radio-frequency detection using high-orbital-angular-momentum states in a {Rydberg}-atom receiver},
    volume = {6},
    doi = {10.1103/PhysRevResearch.6.023317},
    number = {2},
    journal = {Physical Review Research},
    author = {Allinson, Gianluca},
    year = {2024},
}

@book{steck_quantum_nodate,
    edition = {(revision 0.16.6, 17 October 2025)},
    title = {Quantum and {Atom} {Optics}},
    url = {http://steck.us/teaching},
    author = {Steck, Daniel A},
}

@article{holloway_electric_2017,
    title = {Electric field metrology for {SI} traceability: {Systematic} measurement uncertainties in electromagnetically induced transparency in atomic vapor},
    volume = {121},
    issn = {0021-8979, 1089-7550},
    shorttitle = {Electric field metrology for {SI} traceability},
    url = {https://pubs.aip.org/jap/article/121/23/233106/143771/Electric-field-metrology-for-SI-traceability},
    doi = {10.1063/1.4984201},
    number = {23},
    urldate = {2024-09-11},
    journal = {Journal of Applied Physics},
    author = {Holloway, Christopher L. and Simons, Matt T. and Gordon, Joshua A. and Dienstfrey, Andrew and Anderson, David A. and Raithel, Georg},
    month = jun,
    year = {2017},
    pages = {233106},
}

@article{prajapati_enhancement_2021,
    title = {Enhancement of electromagnetically induced transparency based {Rydberg}-atom electrometry through population repumping},
    volume = {119},
    issn = {0003-6951},
    url = {https://doi.org/10.1063/5.0069195},
    doi = {10.1063/5.0069195},
    number = {21},
    urldate = {2024-09-10},
    journal = {Applied Physics Letters},
    author = {Prajapati, Nikunjkumar and Robinson, Amy K. and Berweger, Samuel and Simons, Matthew T. and Artusio-Glimpse, Alexandra B. and Holloway, Christopher L.},
    month = nov,
    year = {2021},
    pages = {214001},
}

@article{rotunno_investigating_2023,
    title = {Investigating electromagnetically induced transparency spectral lineshape distortion due to non-uniform fields in {Rydberg}-atom electrometry},
    volume = {134},
    issn = {0021-8979},
    url = {https://doi.org/10.1063/5.0161213},
    doi = {10.1063/5.0161213},
    number = {8},
    urldate = {2024-07-25},
    journal = {Journal of Applied Physics},
    author = {Rotunno, Andrew P. and Holloway, Christopher L. and Prajapati, Nikunjkumar and Berweger, Samuel and Artusio-Glimpse, Alexandra B. and Brown, Roger and Simons, Matthew and Robinson, Amy K. and Kayim, Baran N. and Viray, Michael A. and Jones, Jasmine F. and Sawyer, Brian C. and Wyllie, Robert and Walker, Thad and Ziolkowski, Richard W. and Jefferts, Steven R. and Geibel, Steven and Wheeler, Jonathan and Imhof, Eric},
    month = aug,
    year = {2023},
    pages = {084401},
}

@article{raab_hf_2002,
    title = {{HF}, {VHF}, and {UHF} systems and technology},
    volume = {50},
    issn = {1557-9670},
    url = {https://ieeexplore.ieee.org/document/989972},
    doi = {10.1109/22.989972},
    number = {3},
    urldate = {2026-05-19},
    journal = {IEEE Transactions on Microwave Theory and Techniques},
    author = {Raab, F.H. and Caverly, R. and Campbell, R. and Eron, M. and Hecht, J.B. and Mediano, A. and Myer, D.P. and Walker, J.L.B.},
    month = mar,
    year = {2002},
    pages = {888--899},
}

@inproceedings{zhang_propagation_2009,
    title = {Propagation comparisons at {VHF} and {UHF} frequencies},
    issn = {2164-2974},
    url = {https://ieeexplore.ieee.org/document/4957352},
    doi = {10.1109/RWS.2009.4957352},
    urldate = {2026-05-19},
    booktitle = {2009 {IEEE} {Radio} and {Wireless} {Symposium}},
    author = {Zhang, Xiaohu and Burress, Thomas Weston and Albers, Keith B. and Kuhn, William B.},
    month = jan,
    year = {2009},
    note = {ISSN: 2164-2974},
    pages = {244--247},
}

@article{mattioni_design_2007,
    title = {Design of a {Broadband} {HF} {Antenna} for {Multimode} {Naval} {Communications}—{Part} {II}: {Extension} to {VHF}/{UHF} {Ranges}},
    volume = {6},
    issn = {1548-5757},
    shorttitle = {Design of a {Broadband} {HF} {Antenna} for {Multimode} {Naval} {Communications}—{Part} {II}},
    url = {https://ieeexplore.ieee.org/document/4137531/references},
    doi = {10.1109/LAWP.2007.893094},
    urldate = {2026-05-19},
    journal = {IEEE Antennas and Wireless Propagation Letters},
    author = {Mattioni, L. and Marrocco, G.},
    year = {2007},
    pages = {83--85},
}

@article{sedlacek_microwave_2012,
    title = {Microwave electrometry with {Rydberg} atoms in a vapour cell using bright atomic resonances},
    volume = {8},
    copyright = {2012 Springer Nature Limited},
    issn = {1745-2481},
    url = {https://www.nature.com/articles/nphys2423},
    doi = {10.1038/nphys2423},
    number = {11},
    urldate = {2024-09-11},
    journal = {Nature Physics},
    publisher = {Nature Publishing Group},
    author = {Sedlacek, Jonathon A. and Schwettmann, Arne and Kübler, Harald and Löw, Robert and Pfau, Tilman and Shaffer, James P.},
    month = nov,
    year = {2012},
    pages = {819--824},
}

@article{sedlacek_atom-based_2013,
    title = {Atom-{Based} {Vector} {Microwave} {Electrometry} {Using} {Rubidium} {Rydberg} {Atoms} in a {Vapor} {Cell}},
    volume = {111},
    url = {https://link.aps.org/doi/10.1103/PhysRevLett.111.063001},
    doi = {10.1103/PhysRevLett.111.063001},
    number = {6},
    urldate = {2024-09-03},
    journal = {Physical Review Letters},
    publisher = {American Physical Society},
    author = {Sedlacek, J. A. and Schwettmann, A. and Kübler, H. and Shaffer, J. P.},
    month = aug,
    year = {2013},
    pages = {063001},
}

@book{stutzman_antenna_2013,
    address = {Hoboken, NJ},
    edition = {3. ed},
    title = {Antenna theory and design},
    isbn = {978-0-470-57664-9},
    publisher = {Wiley},
    author = {Stutzman, Warren L. and Thiele, Gary A.},
    year = {2013},
}

@misc{noauthor_rydberg-state_nodate,
    title = {Rydberg-{State} {Engineering}: {Investigations} of {Tuning} {Schemes} for {Continuous} {Frequency} {Sensing} {\textbar} {Phys}. {Rev}. {Applied}},
    url = {https://journals.aps.org/prapplied/abstract/10.1103/PhysRevApplied.19.044049},
    urldate = {2026-05-19},
}

@article{richardson_extraction_2025,
    title = {Extraction of {Effective} {Electromagnetic} {Material} {Properties} for {Rydberg} {Electrometer} {Vapor} {Cells} from 10-300 {MHz}},
    url = {http://dx.doi.org/10.36227/techrxiv.173950932.23386150/v1},
    doi = {10.36227/techrxiv.173950932.23386150/v1},
    publisher = {Institute of Electrical and Electronics Engineers (IEEE)},
    author = {Richardson, Daniel and Dee, J and Yaeger, J and Viray, M and Marsh, J and Kayim, B and Sawyer, B C and Mantia, D S La and Wyllie, R and Westafer, R S},
    month = feb,
    year = {2025},
}

@misc{kayim_calibration_2026,
    title = {Calibration of electric fields in low-frequency off-resonant {Rydberg} receivers},
    url = {http://arxiv.org/abs/2603.10898},
    doi = {10.48550/arXiv.2603.10898},
    urldate = {2026-05-29},
    publisher = {arXiv},
    author = {Kayim, Baran and Viray, Michael A. and Mantia, David S. La and Richardson, Daniel and Dee, James and Westafer, Ryan S. and Sawyer, Brian C. and Wyllie, Robert},
    month = mar,
    year = {2026},
    note = {arXiv:2603.10898 [physics.atom-ph]},
}

@misc{talashila_floquet_2026,
    title = {Floquet {Interpretation} of {Avoided} {Crossings} in {AC} {Stark}-{Shifted} {Rydberg}-{EIT} {Spectra}},
    url = {https://arxiv.org/abs/2609.06062v1},
    urldate = {2026-09-14},
    journal = {arXiv.org},
    author = {Talashila, Rajavardhan and Prajapati, Nikunjkumar and Schlossberger, Noah and Holloway, Christopher L.},
    month = sep,
    year = {2026},
}

@article{glick_low-frequency_2026,
    title = {Low-frequency electric field sensing with a {Rydberg} beam},
    volume = {114},
    issn = {2469-9926, 2469-9934},
    url = {https://link.aps.org/doi/10.1103/69rr-cf77},
    number = {1},
    urldate = {2026-09-16},
    journal = {Physical Review A},
    author = {Glick, Jeremy and Dickson, John R. and Wood, Josie and Kunz, Paul},
    month = jul,
    year = {2026},
    pages = {012608},
}

\end{document}